%% file: Baryon-Meson-Susy-II.tex
\documentclass[12pt]{article}
\usepackage[colorlinks=true, urlcolor=navyblue, linkcolor=navyblue, citecolor=navyblue]{hyperref}
\usepackage{graphicx,amsmath,amssymb}
\usepackage{indentfirst}
\usepackage[usenames]{color}
\usepackage{epstopdf}
\usepackage{enumerate}
\usepackage{appendix}
\usepackage{setspace}

\definecolor{navyblue}{rgb}{0,0.08,0.45}
\definecolor{darkred}{rgb}{0.7,0.0,0.0}
\definecolor{darkgreen}{rgb}{0,0.6,0.2}

\input{definitions}

\begin{document}


\begin{flushright}
{\small SLAC--PUB--16184 \\ \vspace{2pt}}
\end{flushright}

\vspace{30pt}

\begin{center}
{\huge  { Superconformal Baryon-Meson Symmetry}}
\end{center}

\vspace{-15pt}

\begin{center}
{\huge  { and Light-Front Holographic QCD}}
\end{center}

\vspace{20pt}

\centerline{Hans G\"unter Dosch}

\vspace{3pt}

\centerline{\it Institut f\"ur Theoretische Physik, Philosophenweg
16, D-6900 Heidelberg,
Germany~\footnote{{\href{mailto:dosch@thphys.uni-heidelberg.de}{\tt
h.g.dosch@thphys.uni-heidelberg.de}}}}

\vspace{8pt}

\centerline{Guy F. de T\'eramond}

\vspace{3pt}

\centerline {\it Universidad de Costa Rica, San Jos\'e, Costa
Rica~\footnote{{\href{mailto:gdt@asterix.crnet.cr}{\tt
gdt@asterix.crnet.cr}}}}

\vspace{8pt}

\centerline{Stanley J. Brodsky}

\vspace{3pt}

\centerline {\it SLAC National Accelerator Laboratory, Stanford
University, Stanford, CA 94309,
USA~\footnote{\href{mailto:sjbth@slac.stanford.edu}{\tt
sjbth@slac.stanford.edu}}}

\vspace{40pt}

\begin{abstract}

We construct an effective QCD light-front Hamiltonian for both
mesons and baryons  in the chiral limit based on the generalized
supercharges of a superconformal graded algebra.  The
superconformal construction is shown to be equivalent to a
semi-classical approximation to light-front QCD and its embedding
in AdS space.   The specific breaking of conformal invariance
inside the graded algebra uniquely determines the effective
confinement potential. The generalized supercharges connect the
baryon and meson spectra to each other in a remarkable manner. In
particular, the $\pi/b_1$ Regge trajectory is  identified as the
superpartner of the nucleon trajectory.  However, the lowest-lying
state on this trajectory, the $\pi$-meson is massless in the
chiral limit and has no supersymmetric partner.

\end{abstract}

\newpage

\tableofcontents


\section{Introduction \label{intro}}

Light front holographic QCD (LFHQCD) has brought important
insights into hadron dynamics, especially  to the color
confinement problem. In Refs.
\cite{Brodsky:2006uqa,deTeramond:2008ht} a remarkable equivalence
between the bound-state equations of the light-front Hamiltonian
in 3+1 physical space-time~\footnote{For a review of light-front
physics see~\cite{Brodsky:1997de}.}  and those obtained in
five-dimensional anti-de Sitter space (AdS$_5$) has been observed:
The holographic coordinate $z$ in AdS$_5$ space can be identified
with the  boost-invariant light front (LF) separation $\zeta$
between constituents~\footnote{We therefore will use in the
following the variable $\zeta$ both as light-front variable in
LFHQCD and as the holographic (fifth-dimensional) coordinate in
AdS$_5$.}.   This holographic equivalence allows one to relate the
effective light-front potential for bosons to the
symmetry-breaking factor introduced in AdS$_5$.  In the case of
integer spin fields, the breaking of the conformal isometries of
AdS space can be done by introducing an additional model-dependent
factor into the AdS action -- a dilaton term $e^{\varphi(\zeta)}$.
The specific form of the symmetry-breaking factor, however, is not
fixed {\it a priori}, but it can be deduced from the comparison
with the experimentally observed spectra.  Linear Regge
trajectories demand for $\vp(\zeta)$ the form $\vp(\zeta) =
\la_M\,\zeta^2$ \cite{Karch:2006pv,deTeramond:2010ge}. The
resulting  LF effective potential is harmonic and confining,  and
it also includes a $J$-dependent constant term. This extra term is
a consequence of the separation between kinematical and dynamical
quantities for arbitrary spin~\cite{deTeramond:2013it}, prescribed
by the light-front mapping of AdS bound-state equations. The extra
constant term has important phenomenological consequences; in
particular, it leads in the chiral limit to a massless pion.

A  large step forward in understanding why the effective potential
must have the form of a confining harmonic potential was made by
applying a method developed in conformal quantum mechanics by de
Alfaro, Fubini and Furlan~\cite{deAlfaro:1976je} (dAFF) to the
light-front  bound-state equations~\cite{Brodsky:2013ar}. Starting
from a conformally invariant action, a new Hamiltonian can be
constructed  as a superposition of the generators of  the
conformal algebra. Remarkably, the action remains conformally
invariant, and the form of the resulting confining potential is
uniquely fixed~\cite{Brodsky:2013ar}. It has the form of a
harmonic oscillator and corresponds to the quadratic  dilaton term
previously introduced before by purely phenomenological
arguments~\cite{Karch:2006pv,deTeramond:2010ge}.  However, the
$J$-dependent constant term,  referred to above,  cannot be
derived from the DAFF procedure. Furthermore, for half-integer
spin a dilaton term in the AdS action does not lead to
confinement~\cite{Kirsch:2006he}, and therefore an additional
Yukawa-like interaction term $\bar \psi V(\zeta) \psi$ has to be
added to the fermionic action. This Yukawa interaction term
$V(\zeta)$ again has to be determined phenomenologically --  one
finds that the linear baryon Regge trajectories, with equal
spacing in the orbital and radial excitations, as observed
phenomenologically, requires the form $V(\zeta) = \la_B \,
\zeta$~\cite{Brodsky:2008pg, Abidin:2009hr}.

Recently, we have shown~\cite{deTeramond:2014asa} that a
comparison of the half-integer LF bound-state equations with the
Hamiltonian equations of superconformal quantum mechanics fixes
the form of the LF potential in full agreement with the
phenomenologically deduced form $V(\zeta) = \la_B \, \zeta$. This
procedure, originally developed by  Fubini and Rabinovici
(FR)~\cite{Fubini:1984hf}, is the superconformal extension of the
procedure applied by dAFF~\cite{deAlfaro:1976je}.  In brief: a new
evolution Hamiltonian   can be constructed using a generalized
supercharge which is a superposition of the original supercharge
together with a spinor operator which occurs only in the
superconformal algebra.
 The resulting superconformal quantum mechanics applied to the fermionic  LF bound-state equations  is completely dual to AdS$_5$; this is in contrast to conformal quantum mechanics without supersymmetry, which is dual to the bosonic sector of AdS$_5$ only up to a constant term, which in turn is fixed only by embedding the LF wave equations for arbitrary integer spin into AdS$_5$.

As we shall discuss in this paper, superconformal quantum
mechanics applied to LF Holographic QCD also implies striking
similarities between the meson and baryon spectra.  In fact,
as we shall show,  the holographic QCD light-front Hamiltonians
for the states on the pion and proton trajectories are identical
if one shifts the internal  angular momenta of the meson ($L_M$)
and baryon ($L_B$)  by one unit: $L_M=L_B+1$.   The baryon and
meson trajectories are  actually observed to be linear in the
squared masses $M^2 \propto (n+L)$, as predicted by holographic
QCD,  a feature not obvious for states satisfying effective
bound-state equations (Dirac or generalized Rarita-Schwinger).
The slope of the  trajectories in the principal quantum number $n$
and the orbital angular momentum $L$ are  also very similar.   In
fact, the best fits to the numerical values for the Regge slopes
agree within $\pm 10 \%$ for all hadrons, mesons and baryons; this
leads to a near-degeneracy of meson and baryon levels in the
model.

 The idea to apply supersymmetry to hadron physics is certainly not new~\cite{Miyazawa:1966mfa, Catto:1984wi, Lichtenberg:1999sc}.
 In~\cite{Miyazawa:1966mfa} mesons and baryons are grouped together in a big supermultiplet, a representation of  $U_{6/21}$. In~\cite{Catto:1984wi}
  the supersymmetry results of Miyazawa~\cite{Miyazawa:1966mfa}  are recovered in a QCD framework, provided that a diquark configuration emerges through an effective string interaction. This approach relies heavily on the fact that in $SU(3)_C$ a diquark can be in the same color representation as an antiquark, namely a $\bar 3$. A meson is formed by a quark-antiquark pair and a baryon by a quark and a diquark, which remains color singlet. It is plausible to assume that the color force between a quark and a diquark is approximately equal to that between a quark and an antiquark; and from this, an effective supersymmetry between mesons and baryons follows. An apparent difficulty in this approach is that the pion and the nucleon would have the same mass and thus, supersymmetry would be badly broken~\cite{Lichtenberg:1999sc}.  In fact, in the chiral limit  -- the limit of massless quarks -- the pion is massless, and this state has no obvious supersymmetric partner: there is no (nearly) massless baryonic state. In the direct diquark approach~\cite{Miyazawa:1966mfa, Catto:1984wi, Lichtenberg:1999sc} there is no natural way to take into account the special role of the pion.

In certain aspects,  our approach is similar to the diquark
picture described above.  The light-front  clustering
decomposition used here divides the baryon constituents  into
a special constituent, the active quark, and the rest, the
spectator cluster, which  could be identified with a diquark.  However, in
contrast to the direct diquark picture, the problem of a baryonic
partner of the pion does not occur in our approach.  It  yields a
massless pion, but the supercharge, which  transforms meson into
baryon wave  functions, annihilates the pion wave function and
therefore it has no baryonic partner. The details of
this mechanism, which only occurs for a massless pion, are
explained in Sec. \ref{NT}.

The approach described here, in contrast to the
direct diquark picture of Refs.~\cite{Miyazawa:1966mfa, Catto:1984wi, Lichtenberg:1999sc}, is by no means restricted to a special number
of colors. Indeed, in this effective theory the color  quantum number does not
appear explicitly. However, since it is  an offspring of the Maldacena AdS/CFT
correspondence~\cite{Maldacena:1997re}, it is reminiscent of
an $N_C \to \infty$  theory. This interpretation is also in accordance with
the zero width of all states, including the excited ones.  It is
interesting to note that there exists a genuine supersymmetric
approach to the meson-baryon relation relying on the $N_C \to
\infty$  limit.   Armoni and Patella~\cite{Armoni:2009zq} consider
$\cN=1$ supersymmetric $SU(N_C$);  in their approach,  the meson is formed by a
bosonic string from  a quark to an anti-quark, whereas the baryon
is formed by a fermionic string between two quarks. In the large
$N_C$ limit the string tension for both objects become equal:
``Supersymmetric relics"~\cite{Armoni:2003fb} from the
supersymmetric theory lead to  equal string tension for mesons and
baryons in $SU(N_C$).

We emphasize that the supersymmetric relations between the observed
baryons and mesons, which we derive here, are not  a
consequence of  supersymmetric QCD with scalar quarks and gluinos.
Since no supersymmetric partners of the fundamental QCD fields
have been observed, such a theory is evidently broken below the
TeV scale. The relations derived here are relations between the
wave functions of  hadrons, not field operators. The relations
obtained in the framework of supersymmetric quantum mechanics
reflect properties of the confining mechanism in an effective
semiclassical theory.   One thus expects deviations from
experiment which are of the same order  as in  light-front
holographic QCD.

This article is organized as follows:  After briefly reviewing
some important results of light-front holographic QCD in
Sec.~\ref{LFHQCD},  we discuss in Sec.~\ref{SCA} the construction
of the bound-state Hamiltonian within the superconformal algebra
and the breaking of dilation invariance
following~\cite{Fubini:1984hf}.  The search for the supersymmetric
partners of the baryon trajectories is discussed in Sec.
\ref{BMS}.  A summary of the main results and our conclusions are
presented in Sec. \ref{conclusions}. Some useful formulae for the
derivations presented in this article are given in the appendices.

\section{Light Front Holographic QCD \label{LFHQCD}}

We first briefly review some principal results of light-front
holographic QCD~\footnote{For an extended review see
\cite{Brodsky:2014yha}.}. In holographic QCD an integer-spin
field in AdS$_5$, with a free hadronic field at the
four-dimensional border $\zeta=0$,  is  split into a component
$\Phi_J(\zeta)$, describing the behavior in the bulk, and  a plane
wave with  an integer $J$-spinor describing the Minkowski
space-time behavior (See \cite{Brodsky:2014yha}, Sec. 5.1.1):
\beq
 \Phi_{\nu_1 \cdots \nu_J}(P,\zeta) = \Phi_J(\zeta) e^{i P \cdot x} \ep_{\nu_1 \cdots \nu_J} ({P}).
\enq The four-momentum squared is the mass squared of the hadron
represented by the free field,  $P^2=M^2$.

A Schr\"odinger-like  wave equation~\cite{deTeramond:2008ht,
deTeramond:2013it} follows from the AdS action for arbitrary
integer spin-$J$ modified by a dilaton term $e^{\vp(\zeta)}$: 
\beq
\label{LFSE} \left(-\frac{d^2}{d \zeta^2} + \frac{4L^2-1}{4
\zeta^2} + U(\zeta,J)\right) \phi_J(\zeta)   = 0, 
\enq where we
have factored out the scale  $(1/\zeta)^{J - 3/2}$ and  dilaton
factors from the AdS field $\Phi_J$ by writing $\Phi_J(\zeta)   =
\left(R/\zeta\right)^{J - 3/2 } e^{- \varphi(\zeta)/2} \, \phi_J(\zeta)$.
Equation \req{LFSE} has exactly the form of a LF wave equation for
massless quarks with a LF effective potential $U$ and  LF angular
momentum $L$. The latter is related to the total spin $J$ and the
product of the AdS mass $\mu$ with the AdS  radius  $R$ by \beq
\label{muR} (\mu R)^2 = L^2 -(J-2)^2. \enq The potential $U$ is
related to the dilaton profile by~\cite{deTeramond:2010ge, deTeramond:2013it}
\beq U(\zeta,J)=\frac{1}{2} \vp''(\zeta) +
\frac{1}{4} \vp'(\zeta)^2 +\frac{2 J-3}{2 \zeta}\,\vp'(\zeta).
\enq 
The holographic variable $\zeta$ is identified with the LF
invariant  invariant transverse separation: $\zeta^2=b^2_\perp
u(1-u)$~\cite{Brodsky:2006uqa, deTeramond:2008ht}, where $b_\perp$
is the transverse separation of the constituents and $u$  is the
longitudinal light-front momentum fraction.

In the case of the quadratic dilaton profile $\vp(\zeta)=\la_M
\zeta^2$,  the LF effective potential is $ U(\zeta,J)= \la_M^2
\zeta^2 + 2 \la_M (J-1)$, and the holographic bound-state wave
equation  \req{LFSE} can be written as
 \beq  \label{a5}
 \left(- \frac{d^2}{d\zeta^2} + \la_M^2 \,\zeta^2+ 2\la_M \,(J-1) + \frac{4 \nu^2-1 }{4 \zeta^2}\right)\phi_{J} = M^2\,  \phi_{J},
\enq for  a meson with total spin $J$. Near $\zeta =0$ the regular
solution behaves as $\phi_J(\zeta) \sim \zeta^{\nu+\half}$,
corresponding to twist $2+\nu$. In LFHQCD one thus has   $\nu=
L_M$, where $L_M$  is the LF angular momentum of the meson, $L_M =
\vert L^z_M\vert_{\rm max}$. The  eigenvalues of \req{a5} predict
the meson spectrum \beq \label{meson-spec} M^2_{n, L, J}= 4
\left(n + \frac{J + L_M}{2}\right)\, \la_M , \enq for $\la_M >0$,
where $n$ indicates the radial excitation quantum number:  the
number of notes in the wavefunction.

Similarly, the AdS  field for arbitrary half-integer spin-$J$ can
be factorized into a bulk wave function $\Psi^\pm _J (\zeta)$ and
a plane wave with a Rarita-Schwinger or Dirac spinor with momentum
$P$ and mass $M$, representing a freely propagating  baryon at the
AdS border (See \cite{Brodsky:2014yha}, Sec. (5.2))
\beq
\Psi^\pm_{\nu_1 \cdots \nu_{J-1/2}}(P, \zeta) =  \Psi^\pm _J
(\zeta) \, e^{iPx} u^{\pm}_{\nu_1 \cdots \nu_{J-1/2}}({P}), 
\enq
where the chiral spinors $u^{\pm}_{\nu_1 \cdots \nu_{J-1/2}}=
\half(1\pm \ga_5) u_{\nu_1 \cdots \nu_{J-1/2}}$ satisfy the
equations \beq \ga \cdot P\; u^{\pm}_{\nu_1 \cdots
\nu_{J-1/2}}({P})= M u^{\mp}_{\nu_1 \cdots \nu_{J-1/2}} ({P}) \, ;
\quad \ga_{\nu_1} \,u^{\pm}_{\nu_1 \cdots \nu_{J-1/2}}({P})= 0.
\enq The spinors $u^{\pm}$ have positive and negative chirality,
respectively.

The bound-state wave  equations for the AdS bulk wave functions
$\Psi_J^\pm$ can be derived from the action for arbitrary
half-integer spin-$J$ if one includes  the effective interaction
$V(\zeta)= \la_B \zeta$. The result is~\cite{deTeramond:2013it}
 \beqa
 \left(-\frac{d^2}{d \zeta^2} + \la_B^2\, \zeta^2 + 2 \la_B \,(\nu +\half)  +\la_B  + \frac{4\nu^2 -1}{4 \zeta^2}\right)\psi_J^+ &  \! = \! & M^2 \, \psi_J^+,
\label{a3}\\
 \left(- \frac{d^2}{d \zeta^2} + \la_B^2 \,\zeta^2+ 2 \la_B \,(\nu +\half)  -\la_B + \frac{4 (\nu +1)^2-1 }{4 \zeta^2}\right)\psi_J^- & \! = \!  &M^2 \, \psi_J^-,
 \label{a4}
 \enqa
where we have factored out the scale  $(1/\zeta)^{J - 5/2}$ by writing
$\Psi^\pm_J(\zeta)   = \left(R/\zeta\right)^{J - 5/2 }   \psi^\pm_J(\zeta)$,
and $\nu$ is related to the product of the AdS fermionic mass and
the AdS radius $R$ by \beq \nu = \mu R - \half . \enq The baryon
spectrum which follows from (\ref{a3},\ref{a4}) is \beq
\label{baryon-spec} M^2_{n,\nu} = 4( n + \nu + 1)\, \la_B , \enq
for $\la_B >0$.  The eigenvalues given by \req{baryon-spec} do not
depend explicitly on $J$, an important  result also found in
Ref.~\cite{Gutsche:2011vb}.

\section{Superconformal Algebra and Breaking of Dilatation Symmetry \label{SCA}}

We will now show how the preceding results can be systematically
derived using superconformal algebra, but with important new
consequences. One starts with the simplest graded algebra of two
fermionic operators, the supercharges  $Q$ and $Q^\dagger$,  and a
Hamiltonian $H$ \cite{Witten:1981nf} \beqa \label{sysy-al} && \{Q,
Q^\dagger\}= 2 H, \\ &&\{Q,Q\} = \{Q^\dagger,Q^\dagger\}=0, \\
&&[Q,H]=[Q^\dagger,H]=0. \enqa A simple realization is: \beq Q=
\psi^\dagger(-i p + W), \quad \quad Q^\dagger= \psi(i p + W), \enq
where $p$ is the canonical momentum operator; $\psi$ and
$\psi^\dagger$ are fermionic operators with anti commutation
relation \beq \{\psi,\psi^\dagger\}=1, \enq and $W$ is an
arbitrary potential (the superpotential).

A realization using  Pauli matrices $\vec \si $  is: \beq \psi=
\half(\si_1-i\si_2), \quad \quad  \psi^\dagger=
\half(\si_1+i\si_2), \enq
 leading to
 \beq \label{B}
 B=\half  [\psi^\dagger,\psi]=\half \si_3,
\enq where $B$ is the generator of $U(1)$ transformations $\psi
\to e^{i \al} \psi$,  $\psi^\dagger  \to e^{- i \al} \psi^\dagger$
with eigenvalues $+\half$ and $-\half$.

In the Schr\"odinger picture the supercharges  are realized as
operators in $\cL_2(R_1)$, with $ p = - i \, d/dx$: \beq Q =
\psi^\dagger\left( - \frac{d}{dx} + W(x)\right), \enq and \beq
Q^\dagger= \psi \left(\frac{d}{dx} +W(x) \right), \enq leading to
the supersymmetric Hamiltonian: \beq \label{Ham} H = \half\{Q,
Q^\dagger\} = \frac{1}{2} \left(- \frac{d^2}{d x^2}  + W^2(x) -  2
W'(x)\, B \right). \enq The Hamiltonian operates on 2-spinors \beq
\label{phi} \vert \phi\rangle =
\begin{pmatrix}
\phi_1\\  \phi_2
\end{pmatrix} ,
\enq of which one component can be attributed  to fermion number 1
and the other 0.  Imposing  conformal symmetry  leads  to an
unique choice of $W$ \cite{Akulov:1984uh, Fubini:1984hf}, namely
\beq \label{conf-pot} W(x) = \frac{f}{x} , \enq with a
dimensionless constant $f$.

Introducing the spinor operators 
\beq 
S=  \psi^\dagger \, x, \quad \quad S^\dagger = \psi \, x ,
\enq
 one can construct the larger
graded algebra \cite{Haag:1974qh} (superconformal algebra), which
contains the conformal algebra with the dilatation generator $D$
and the special conformal transformation generator $K$. The
extended algebraic structure is 
\beqa \label{susy-extended}
\half\{Q,Q^\dagger\} &=& H, \quad \quad \half\{S,S^\dagger\}=K, \\
\{Q,S^\dagger\} &=& f - B + 2 i D, \\
 \{Q^\dagger,S\} &=& f - B - 2 i D,
\enqa 
where the operators \beqa
H & = & \frac{1}{2} \left( - \frac{d^2}{d x^2}  + \frac{f^2 + 2 B f}{x^2} \right), \\
 K  & = & \half x^2, \\
 D & = & \frac{i}{4} \left(\frac{d}{dx} x + x \frac{d}{d x} \right),
\enqa 
satisfy the conformal algebra \beq [H,D]= i H, \quad\quad
[H,K]= 2 i D, \quad \quad [K,D]=-i K . \enq The anti-commutators
of all the other generators vanish:  $\{Q,Q\} =  \{Q,S\} = \cdots
= 0$.

Fubini and Rabinovici considered several ways to construct a new
compact quantum mechanical evolution operator inside the
superconformal algebra. The most straightforward way is to
directly follow the procedure of dAFF~\cite{deAlfaro:1976je} and
construct  a linear combination of the (old) Hamiltonian and the
generator of special conformal
transformations~\cite{Akulov:1984uh,Fubini:1984hf}. There is,
however, the interesting possibility of constructing a new
Hamiltonian using the superposition of  generalized
supercharges within the extended graded algebra~\cite{Fubini:1984hf} and thus preserving
supersymmetry. This is the procedure we shall follow here. To this
end, we slightly generalize the definitions of
FR~\cite{Fubini:1984hf} and introduce a new supercharge $R$ as a
linear combination of the generators $Q$ and $S$ \beq
\label{super-R} R_{\la} =   Q + \la \, S. \enq This leads, in
analogy to the dAFF procedure in conformal quantum mechanics, to
the introduction of a constant with nonzero dimensions; in fact,
since $Q$ has dimension $[x^{-1}]$, and $S$  has dimension
$[x^1]$, $\la$ must therefore have dimension $[x^{-2}]$ .

One can now construct  a new  evolution operator $G$ inside the
superconformal algebra in terms of the new supercharge $R$: 
\beqa
\label{Rsusy1}
&& \{R_\la, R_\la^\dagger\} = G , \\ \label{Rsusy2}
&&\{R_\la,R_\la\} = \{R_\la^\dagger,R_\la^\dagger\}=0, \\ \label{Rsusy3}
&&[R_\la,G]=[R_\la^\dagger,G]=0. \enqa We find \beq \label{Gwu} G
= 2   H + 2 \la^2  K +  2 \la \,( f \mbox{\bf I} - B), \enq which
is a compact operator for $\la \in \cal{R}$.

The supercharge operator $R_\la^\dagger$ transforms a state $\vert
\phi \rangle$  into the state  $R^\dagger  \vert \phi \rangle$
with different fermion number (See Appendix \ref{Oq}). By
construction,  the evolution operator $G$ commutes with $R_\la$;
it thus follows that the states  $\vert \phi \rangle$ and
$R^\dagger  \vert \phi \rangle$ have identical eigenvalues. In
fact, if $ \vert \phi_E\rangle$ is an eigenstate of $G$ with
$E\neq 0$, \beq  \label{EE} G\,|\phi_E\rangle = E
\,|\phi_E\rangle, \enq then $G \,R^\dagger_\la  \,  |\phi_E\rangle
= R^\dagger_\la \,G\, |\phi_E\rangle=E\,  R^\dagger_\la
|\phi_E\rangle$,  and thus  $R^\dagger_\la  \,  |\phi_E\rangle$ is
also an eigenstate  of $G$ with the same eigenvalue.

The new Hamiltonian  $G$ is diagonal. In the Schr\"odinger
representation: \beqa \label{a1}
G_{11}&=&\left(- \frac{d^2}{d x^2} + \la^2 \,x^2+ 2 \la\, f  - \la+ \frac{4 (f+\half)^2 - 1}{4 x^2}\right), \\
\label{a2} G_{22}&=&\left(- \frac{d^2}{d x^2} + \la^2 \, x^2 + 2
\la\, f  + \la + \frac{4 (f-\half)^2 -1}{4 x^2}\right), \enqa with
$ G_{10}= G_{01}=0. $ For $f \geq \half$ and $ \la > 0$  the
spectra of both operators are identical: \beq \label{susy-spec}
 E_n= 4 \left(n + f  + \half \right) \la.
\enq

Comparing (\ref{a3}, \ref{a4}) with (\ref{a1}, \ref{a2}) we
recover the result of Ref.~\cite{deTeramond:2014asa}, namely that
the modified Hamiltonian G of superconformal quantum mechanics  is
the same as the Hamiltonian derived in LF holographic QCD,
provided we identify $\phi_2(x)$, the eigenfunction of $G_{22}$,
with the positive  chirality wave function $\psi_J^+(\zeta)$,
identify $\phi_1(x)$, the eigenfunction of $G_{11}$,  with
$\psi_J^- (\zeta)$; and take $f-\half = \nu = L_B$ and
$\la=\la_B$. The consequences of this remarkable result have been
discussed extensively in Ref.~\cite{deTeramond:2014asa}.

{ In Ref.~\cite{deTeramond:2014asa} the $U(1)$ operator \req{B} $B
= [\psi^\dagger, \psi]$ was identified in the light-front with the
Dirac matrix $\ga_5$ which acts on physical spinors.  In that
paper we  showed that the supercharges relate the chirality-plus
component of a baryonic wave function with the chirality-minus
component of the same baryonic state.  In the usual applications
of supersymmetry, however, the supercharges connect bosonic to
fermionic states. We therefore shall explore in the next section
the possibility to relate mesonic with baryonic wave functions by
the supercharges within the extended graded algebra. 
In this case, the supercharges act on some internal
space. The supercharges in~\cite{deTeramond:2014asa} and those
used  in the following are therefore only formally related. The
bosonic operators $H, D$ and $K$, however, have in both cases the
same physical meaning.}    In particular, we will show that the
$G_{11}$ and $G_{22}$ equations \req{a1} and \req{a2}  match our
light-front holographic equations for both the pion  and nucleon
trajectories.   The extension of this superconformal connection to
the $\Delta$-$\rho$ families will also be discussed.

\section{Baryon-Meson Supersymmetry \label{BMS}}

\subsection{The Superpartner of the Nucleon Trajectory \label{NT}}

In the case of baryons, the assignment of the leading-twist
parameter $\nu$ in Eqs. (\ref{a3}, \ref{a4}), as given in Table
\ref{nuT}~\cite{deTeramond:2014asa}, successfully describes the
structure of  the light baryon orbital and radial
excitations~\footnote{The `leading-twist' assignment referred to
here is the effective twist of the baryonic quark-cluster system;
it  is thus equal to two.  This is in distinction to the usual
application of  twist for hard exclusive  processes which emerges
when the baryon cluster is resolved at high momentum transfer and
is thus equal to the total number of components.}The assignment
$\nu=L_B$ for the lowest trajectory, the nucleon trajectory,  is
straightforward and follows from the stability of the ground state
-- the proton -- and the mapping to LF quantized QCD.

The bound-state equations for the nucleon trajectory are (cf. Eqs.
(\ref{a3}, \ref{a4})): \beqa
 \left(-\frac{d^2}{d\zeta^2} + \la_B^2\, \zeta^2 + 2 \la_B (L_B +1)   + \frac{4 L_B^2 -1}{4 \zeta^2}\right)\psi_J^+ &=& M^2 \, \psi_J^+ , \label{a3n}\\
 \left( -\frac{d^2}{d\zeta^2} + \la_B^2 \,\zeta^2+ 2 \la_B \, L_B  + \frac{4 (L_B+1)^2-1 }{4 \zeta^2}\right)\psi_J^- &=&M^2 \, \psi_J^- . \label{a4n}
\enqa

\begin{table}
\caption{\label{nuT} \small Orbital quantum number assignment for
the leading-twist parameter $\nu$ for baryon trajectories
according to parity $P$ and internal spin $S$.}
\begin{center}
\begin{tabular}{ l | c r }
\hline\hline
 & $~~S = \half$ & \, $~~~~ S = \threehalf$ \ ~~\\ [0.0ex]
 \hline
P = + ~& $~~\nu = L_B$ &  $ \nu = L_B+\half$ \\ [-0.0ex]
 P = \ -- ~ &  $~ \nu = L_B+\half $ &  $\nu = L_B+ 1$ \\ [0.0ex]
 \hline\hline
\end{tabular}
\end{center}
\end{table}

We will now search for the meson supersymmetric partners of the
nucleon trajectory. We choose as starting point  the leading-twist
chirality component $\psi_J^+(\zeta)$ which satisfies \req{a3n}.
With the identifications  $x= \zeta, \; f-\half =L_B$ and
{$\la=\la_B$}, the plus chirality component $\psi_J^+(\zeta)$  is
also an eigenfunction of $G_{22}$, Eq. \req{a2}.  This
identification allow us to define an  effective ``baryon
number'' $N_B$ as a convenient convention to label our ``meson''
and ``baryon" states.  In terms of  the $U(1)$ operator $B =
\half[ \psi^\dagger,\psi]$ 
\beq \label{NB} N_B = \half - B, 
\enq
with eigenvalues 
\beqa
N_B \vert \phi \rangle_M &=& 0 ,\\
N_B \vert \phi\rangle_B &=&  \vert \phi \rangle_B,
\enqa
where $\vert \phi \rangle_B$ has only a lower component ($\phi_1=0$) and $\vert \phi \rangle_M$ only an upper component ($\phi_2=0$):
\beq
\vert \phi\rangle_B  =  \left(\begin{array}{c} 0\\ \phi_2 \end{array} \right),  \quad \quad
\vert \phi\rangle_M =  \left(\begin{array}{c} \phi_1\\ 0 \end{array} \right).
\enq

Therefore, the supersymmetric partner $G_{11}$ \req{a1} should
describe a meson trajectory. Indeed, the Hamiltonian  $G_{11}$
with the above mentioned substitutions agrees with the bound-state
equation \req{a5} for mesons with $J = L_M$, provided we identify
$f+\half=L_M=L_B+1$ and set $\la_M=\la_B$. The lowest state on the
mesonic trajectory, with $J=L_M = 0$ ~~~ -- the pion --   is  massless
in the chiral limit. It corresponds to a negative value of $f$,
namely $f=-\half$ and thus its baryonic partner would have
$L_B=-1$, which is an unphysical state.   As discussed in Appendix
\ref{Op}, this remarkable result, also follows directly from the
superconformal algebra. As  shown there, the operator which
transforms a mesonic state into its baryonic supersymmetric
counterpart, annihilates the meson state if $f=- \half$.

We have thus derived the astonishing result that the pion has no
supersymmetric partner even though no explicit breaking of
supersymmetry has been introduced. Since  the supercharges $R_w, \,
R_w^\dagger$, which connect mesonic and baryonic wave functions,
commute with the Hamiltonian $G$ (\ref{Rsusy1} - \ref{Rsusy3}), it follows that if $\vert
\phi\rangle_M$ is a mesonic state with eigenvalue $E$, then there exists  also a
baryonic state $R_w^\dagger \vert \phi\rangle_M=\vert
\phi\rangle_B$ with the same eigenvalue $E$.  
Indeed
\beq
G\,\vert \phi\rangle_B=G\,R_w^\dagger \vert \phi\rangle_M=R_w^\dagger\,G\, \vert \phi\rangle_M=E\,\vert \phi\rangle_B.
\enq
However, for the specific
eigenvalue $E=0$ we can have the trivial solution 
\beq 
\vert\phi (E=0) \rangle_B =
\left(\begin{array}{c} 0\\0 \end{array} \right).
\enq
This remarkable feature underlines the special role played by the pion in light-front holographic QCD.
As a unique state of zero energy, it plays the same role as the unique vacuum state in a supersymmetric quantum field theory~\cite{Witten:1981nf}~\footnote{In our assignment the Witten index~\cite{Witten:1982df}  for $f=-\half,\; \la>0$ has the value $+1$. It has the same value for $f=\half,\; \la<0$~\cite{Fubini:1984hf}.}.

In is interesting  to note that the  case of negative $f$ was
not considered in~\cite{Fubini:1984hf}, since the classical
potential $ \frac{f}{2 x^2}+ \la^2 x^2$ has no stable ground
state for $f < 0$. Nevertheless, the lowest lying bound state of
$G_{11}$ with $f=-\half$ has the normalizable wave function
$x^{\half} e^{-\la_M x^2/2}$.  This situation is reminiscent of
the AdS/QCD correspondence: angular momentum $L=0$ corresponds to
a tachyonic AdS mass  $\mu^2<0$  (See Eq. \req{muR}),  but
nonetheless the Breitenlohner-Freedman stability bound
\cite{Breitenlohner:1982jf} is still satisfied.

We thus obtain from superconformal quantum mechanics a very
satisfactory result:  both the nucleon and the $I=1, S=0$ mesons
lie on linear trajectories with the same slope and the same radial
and orbital excitation energies.   The lowest lying state on the
meson trajectory is the massless pion. In superconformal quantum
mechanics it corresponds to the value $f=-1/2$, and therefore it
has no supersymmetric partner.

\begin{figure}[h]
\begin{center}
\includegraphics[width=8.6cm]{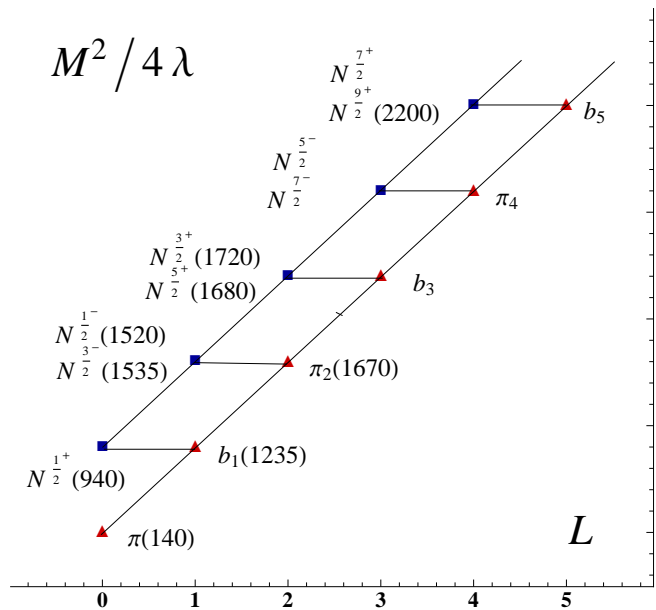}
\end{center}
\caption{\label{MNSCC} \small Meson-nucleon superconformal
connection.  The predicted value of $M^2$  in units of $4 \la$ for
mesons with $S=0$ (red triangles), and baryons with $S=
\frac{1}{2}$ (blue squares) is plotted vs the orbital angular
momentum $L$. The $\pi$-meson has no baryonic partner. The baryon
quantum number assignment is taken from
Ref.~\cite{deTeramond:2014asa}. Nucleon trajectories for $J^z =
L^z \pm S^z$ are degenerate.}
\end{figure}

In the framework  of superconformal quantum mechanics all
eigenstates with eigenvalues different from zero have
supersymmetric partners.  We emphasize that the pion with
$f=-\half$ and zero mass is unique: it  is annihilated by the
fermion-number changing supercharge $R^\dagger_\la$, and it
therefore  has no supersymmetric partner (See Fig.  \ref{MNSCC}).
This is in accordance with the spectroscopy derived from
light-front holographic QCD, where the partners have the masses
$M^2_B=4 \la_B(n+L_B+1)$ and $M^2_M=4 \la_M(n+L_M)$
respectively~\footnote{This result follows from \req{baryon-spec}
and \req{meson-spec}  with $\nu = L_B$ and $J = L_M$,
respectively.}. If one takes $\la_B = \la_M$ in LF holographic
QCD, which is automatic in the superconformal theory, the spectral
results are then identical for $L_M = L_B + 1$.

The predictions of supersymmetric quantum mechanics are based on
the fact that the supercharge operator $R_\la$ transforms baryon
states with angular momentum $L_B$  into their mesonic
superpartners with angular momentum $L_M = L_B + 1$. The operator
$R^\dagger_\la$ operates in the opposite direction.  We thus have
a complete correspondence between light-front holographic QCD and
supersymmetric quantum mechanics. The pion has a very special
role: its  existence is predicted by the superconformal algebra,
and according to the formalism, it is massless and  has no
supersymmetric partner.

The superconformal predictions presented in Fig. \ref{MNSCC}
should be understood as a zeroth-order approximation.  There are,
however, several phenomenological corrections to this initial
approximation. First, the slope of the $\pi/b_1$ trajectory is not
exactly identical to the slope of the nucleon trajectory: for the
mesons $\sqrt{\la_M}= 0.59$ GeV, whereas for the nucleons
$\sqrt{\la_B}= 0.49$ GeV~\cite{Brodsky:2014yha}. This makes the
$b_1$ heavier than its supersymmetric partner, the nucleon. In
terms of LFHQCD this indicates that for this internal spin
configuration, the confining force between the spectator and the
cluster in the baryon is weaker than between the constituents of
the meson;  this makes the meson a more compact object since
$\langle r^2 \rangle \sim 1 /\la$.  Second, the negative parity
nucleon states are systematically higher than the nucleons with
positive parity, a fact which in LF holographic QCD has been taken
into account phenomenologically by the half-integer twist
assignment $\nu = L+\half$ given in Table \ref{nuT}.  It is
expected that this effect could be explained by the different
quark configurations and symmetries of the baryon wave
function~\cite{Klempt:2009pi, Forkel:2008un,
Selem:2006nd}~\footnote{In Ref.~\cite{Selem:2006nd} it is
suggested that this parity level-shift effect in baryons could be
a consequence of the tunneling of the spectator quark into the
cluster.}.

\begin{figure}[h]
\begin{center}
\includegraphics[width=8.6cm]{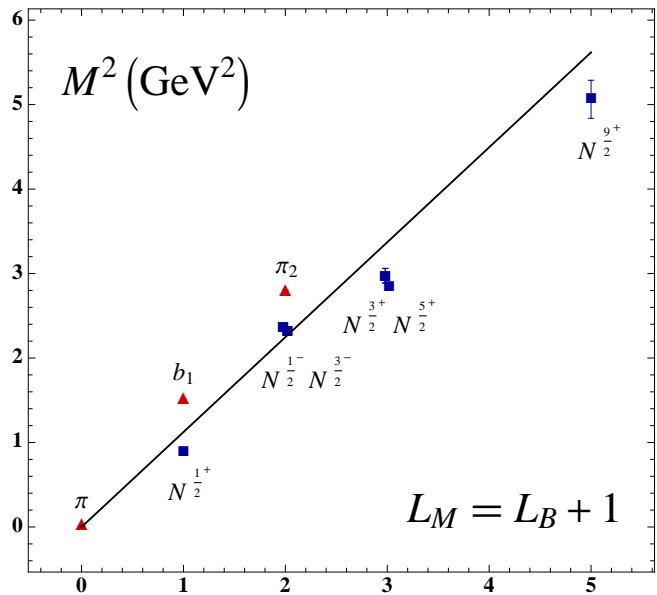}
\end{center}
\caption{\label{fig-pi-n}   \small Supersymmetric meson-nucleon
partners:  Mesons with $ S=0$ (red triangles) and baryons with $S=
\frac{1}{2}$ (blue squares). The experimental values of $M^2$
are plotted vs $L_M = L_B+1$. The solid line corresponds to
$\sqrt \la= 0.53$  GeV.  The $\pi$ has no baryonic partner.}
\end{figure}

The nucleon-meson superpartner pairs are plotted in Fig.
\ref{fig-pi-n} with their measured masses. The observed difference
in the squared masses of the supersymmetric partners indicates
that the most important breaking of supersymmetry is due to the
difference between $\la_B$ and $\la_M$. Only confirmed PDG states
have been included~\cite{Agashe:2014kda}.

\subsection{The Mesonic Superpartners of the Delta Trajectory}

The essential physics derived from the superconformal connection
of nucleons and mesons follows from the action of the
fermion-number changing supercharge operator $R_\la$. As we have
discussed in the previous section, this operator transforms a
baryon  wave function with angular momentum $L_B$  into a
superpartner meson  wave function  with angular momentum
$L_M = L_B + 1$ (See Appendix \ref{Oq}), a state with the
identical eigenvalue -- the hadronic mass squared.  We now check
if this relation holds empirically for other baryon trajectories.

We first observe that baryons with positive parity and internal
spin  $S=\threehalf$, such as the $\Delta^{\threehalf^ +}(1232)$,
and baryons with with negative parity and internal spin $S=\half$,
such as the $\Delta^{\half^ -}(1620)$,  lie on the same
trajectory; this corresponds to  the phenomenological assignment
$\nu =L_B+\half$, given in Table~\ref{nuT}. From \req{baryon-spec}
we obtain the spectrum~\footnote{For the $\Delta$-states this
assignment agrees with the results of Ref.~\cite{Forkel:2007cm}.}
\beq M^{ 2 (+)}_{n, L_B, S = \threehalf} = M^{ 2 (-)}_{n, L_B, S =
\half} = 4 \left(n+L_B + \frac{3}{2} \right) \la_B. \enq If we now
apply the superconformal relation  $L_M = L_B + 1$ and $\la_M =
\la_B$ we predict a meson trajectory with eigenvalues \beq M^2_{n,
L_M} = 4 \left(n+L_M+\frac{1}{2}\right) \la_M, \enq which is,
precisely, the expression for the spectrum of the $\rho$-meson
\req{meson-spec}  for   $J=L_M+1$.   Again, one sees that  the
lowest-lying mesonic state, in this case the $\rho$ meson, has no
superpartner, since $L_M$ would be negative.

\begin{figure}
\begin{center}
\includegraphics[width=8.6cm]{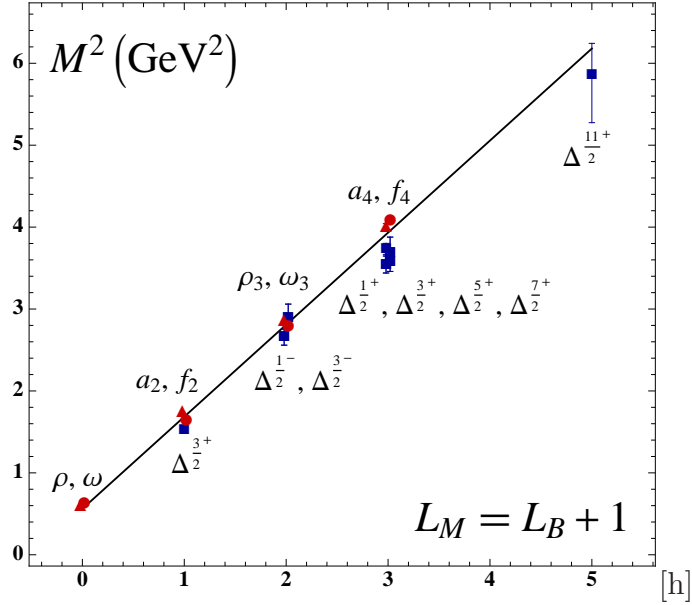}[h]
\end{center}
\caption{\label{rho-delta} \small   Supersymmetric vector-meson
and $\Delta$  partners: Mesons with $ S = 1 $ (red triangles) and
$S = 0$ (red circles) and $\Delta$ states with $S= \frac{3}{2}$
and $S= \frac{1}{2}$ (blue squares) for plus and minus parity
respectively. The experimental values of $M^2$   are plotted vs
$L_M = L_B+1$. The solid line corresponds to  $\sqrt \la= 0.53$
GeV. The $\rho$ and $\omega$ have no baryonic partner, since it
would imply a negative value of $L_B$.}
\end{figure}

Since the phenomenological value of $\la$ for the $\Delta$
trajectory is close to that of the $\rho$ trajectory,
$\sqrt{\la_\Delta}=0.51$ and $ \sqrt{\la_\rho}=0.54$ (See
Ref.~\cite{Brodsky:2014yha}), one can expect good agreement for
the masses of the supersymmetric partners. This is indeed the
case, as can be seen from Fig. \ref{rho-delta}, where we have
included the confirmed $\Delta$ and  $J = L + S$, $S = 1$,
vector-meson states from Ref.~\cite{Agashe:2014kda}.

Using the assignment  $\nu =L_B+\half$  from (Table~\ref{nuT}) and
the comparison of Eqs. \req{a3} with \req{a2} (or \req{a4} with
\req{a1}), we obtain the relation $f = \nu + \half = L_B + 1 =
L_M$ for the superconformal relation $L_M = L_B + 1$. Thus from
\req{a1} we obtain the LF-Hamiltonian for the superpartner vector
meson trajectory \beq \label{a7} G_{11} = \left(- \frac{d^2}{d
\zeta^2} + \la_M ^2 \, \zeta^2+ 2 \la_M (\, L_M - 1)+ \frac{4 (L_M
+ \half)^2 - 1}{4 \zeta^2}\right), \enq with $\la = \la_M =
\la_B$.  {This expression is to be compared with the light-from
holographic Hamiltonian which follows from \req{a5} for $J = L_M +
1$ and $\nu = L_M$: \beq H_{LF} = \left(- \frac{d^2}{d\zeta^2} +
\la_M^2 \,\zeta^2+ 2\la_M \, L_M + \frac{4 L_M^2-1 }{4
\zeta^2}\right).
 \enq }

Thus, by extending the meson-baryon connection for baryons with
$\nu = L_B + \half$  we obtain an identical expression for the
vector meson-spectrum, but  with a different LF Hamiltonian.
This somewhat less satisfactory feature of the $\Delta$-$\rho$
relations is reflected in the transformation under the supercharge
$R_w^\dagger$ (Appendix \ref{Oq}). The  $\rho$ meson wave function
$\phi_1$, that is the  eigenfunction of $G_{11}$ with $f=0$,  is
not annihilated by the action of $R_w^\dagger$ \req{rho-shift}.
Indeed the two Hamiltonians $G_{11}$ and $G_{22}$, \req{a1} and
\req{a2} respectively, are identical for $f=0$. Thus in this case,
the unphysical value of the angular momentum, $L_B=-1$, is  the
only reason to exclude the baryonic  superpartner of the $\rho$.
This is in contrast to the case of the pion, where the
fermion-number changing operator $R_w^\dagger$ actually
annihilates the pion wave function  \req{pi-shift},  since it is a zero mass eigenmode.

\section{Summary and Conclusions \label{conclusions}}

Conformal and superconformal Quantum Mechanics, together with
light-front holographic QCD, has revealed the importance of
conformal symmetry and its breaking for understanding the
confinement mechanism of QCD.

If one introduces the mass scale scale for hadrons using the
method developed by de Alfaro, Fubini and
Furlan~\cite{deAlfaro:1976je}, one obtains a confining theory for
mesons while retaining a conformally invariant action. If one
applies the DAFF procedure to light-front Hamiltonian theory, the
form of the LF  potential is uniquely fixed to that of a harmonic
oscillator in the invariant LF radial variable
$\zeta$~\cite{Brodsky:2013ar}.  It predicts color confinement and
linear Regge meson trajectories with the same slope in the radial
and orbital excitations $n$ and $L$.    If one compares the
construction of the confining LF potential with the Hamiltonian
obtained in light-front holographic QCD, then the dilaton factor in the
AdS action is  uniquely fixed~\cite{Karch:2006pv,
deTeramond:2010ge}. The appearance of the  extra spin-dependent
constant term in the LF potential is a consequence of the specific
embedding of the LF wave equations in AdS for arbitrary
integer-spin~\cite{deTeramond:2013it}. This extra term is
essential for agreement with experiment, including  the prediction
of a  massless pion in the chiral  limit.

In the case of half-integer spin,  the dilaton in the AdS action
does not lead to confinement for baryons since such a term can be
absorbed into the wave function. Confinement thus requires the
addition of a Yukawa-like term in the half-integer spin
Lagrangian.  However, this apparent deficiency of the AdS theory
is cured~\cite{deTeramond:2014asa} by the application of
superconformal quantum mechanics.

Superconformal quantum mechanics can be constructed by restricting
the superpotential in Witten's construction~\cite{Witten:1981nf}
to a conformally invariant
expression~\cite{Fubini:1984hf,Akulov:1984uh}.  Remarkably, it is
possible to introduce a mass scale into the quantum mechanical
evolution equations, without violating
supersymmetry, by introducing a new supercharge which is a linear
combination of generators of the super conformal algebra~\cite{Fubini:1984hf}.   Furthermore, by connecting
the resulting wave equations to the light-front holographic
formalism, one fixes  not only the confining term for baryons and
mesons for all spins, but also the constant terms in the LF
potential. The resulting spectra reproduces  the principal
observed features of mesonic and baryonic Regge trajectories: the
resulting trajectories  are linear, and the spacing of the radial
excitations equals the spacing of the orbital ones. Furthermore,
the baryon masses depend only on the LF angular momentum $L$, but
not on the total spin $J$, as observed in experiment.

There are striking phenomenological similarities between the
baryon and meson spectra which would not be expected from the
underlying quark degrees of freedom, given that in QCD the valence
state in the meson case consists of  confined $q\bar q$
excitations, and baryons are normally considered $qqq$ bound
states. However, the observed Regge trajectories are linear in the
squared mass for both cases, with equal spacings of the orbital and
angular excitations  -- both features which are typical for the
proto-string theory such as the Veneziano
model~\cite{Veneziano:1968yb}. These essential features also follow from the light-front clustering 
properties of the semiclassical approximation to strongly coupled QCD and its holographic
embedding in AdS space. In this approximation a nucleon is effectively a 
quark-diquark object, and it is also described by a one-dimensional effective theory. Furthermore, the coefficients of
the confining term for mesons and baryons agree within  $\pm 10
\,\%$, although they would seem to be completely unrelated.  These
similarities  suggest that supersymmetric relations are
responsible for these remarkable features.

{ In Ref.~\cite{deTeramond:2014asa}  superconformal quantum
mechanics was used to describe baryonic states. There, the
supercharges were shown to relate the positive and negative
chirality components of the baryon wave functions, consistent with
parity conservation. In this paper we have shown that
supercharges, constructed formally as
in~\cite{deTeramond:2014asa}, can also be used to relate hadronic
states with different fermion number. This leads to remarkable
relations between the spectroscopy of baryons and mesons, thus
extending the applicability of light-front superconformal quantum
mechanics to hadronic physics.

An important feature of the Hamiltonian operators (\ref{a1},
\ref{a2}), which act  on the two components of a supermultiplet
$|\phi\rangle$, is the difference in the singular term of the
potential. For one component of the Hamiltonian, it is $\frac{1}{4
x^2}\left( (f+\half)^2 - 1\right)$; for the other component,
it is $\frac{1}{4 x^2}\left( (f-\half)^2 - 1\right)$.  This has
the consequence that the power behavior of the wave function at
the origin (twist) differs by one unit for the two components. In
 in light-front holographic QCD this implies a difference of the LF angular momentum by one
unit $L_M=L_B+1$.  Comparing the spectra of the nucleon and the
$\pi/b_1$  trajectory  one  indeed observes this
degeneracy (See  Fig. \ref{fig-pi-n}).  The leading-twist wave
function of the baryons is identified with the component $\phi_2$
of the supermultiplet $\vert \phi\rangle$, and the wave function
of the mesons is identified with the component $\phi_1$.  As a
consequence, the shared symmetric features of mesons and baryons
are  in fact a consequence of the properties of the superconformal
algebra.

The problem for supersymmetry posed by the pion, which is massless
in the chiral limit, and therefore can have no baryonic
superpartner, is solved in a simple way: The value of the
dimensionless constant $f$ of the conformal potential
\req{conf-pot} has the value $f= L_M-\half = -\half$. The
supercharge $R_\la^\dagger$, \req{super-R}, which transforms the
meson into the baryonic partner, annihilates the pion state, and
therefore there cannot be a baryonic partner. The case $f=-\half$
was not considered by Fubini and Rabinovici~\cite{Fubini:1984hf},
since the classical potential in this case has no lower limit.
Nevertheless, the pion  wave function is regular at the origin and 
normalizable.

We have previously demonstrated a  correspondence between superconformal
quantum mechanics and light-front holographic QCD; however, this demonstration requires both a dilaton term $e^{\la
\zeta^2}$
in the bosonic AdS action,  as well as a Yukawa-like  interaction
term $\la \bar  \psi \, z \, \psi $.  One must also assume in
LFHQCD the same positive value of $\la$ in both terms.  In
contrast, in the superconformal  theory, the equality of    of
$\la$ for mesons and baryon is exact.  (Phenomenologically, this
relation is broken since $\sqrt{\la} = 0.59$ GeV  for the
$\pi/b_1$ and  $\sqrt{\la} = 0.49$ GeV for the nucleon trajectory
(See Fig. \ref{fig-pi-n}).

We have also applied the same procedure  to the $\rho/a_2$ and the
$\Delta$-trajectories.  The wave functions of the
$\rho$-trajectory are identified with the component $\phi_1$,  and
the component $\phi_2$ of the super multiplet is identified with
the $\Delta$-states. As for the case of the $\pi$-nucleon
connection, the properties of the fermion-changing supercharge
$R_\la$ imply that the meson angular momentum $L_M$ is one unit
larger than the baryon angular momentum $L_B$, $L_M = L_B + 1$
consistent with the Hamiltonians (\ref{a1}, \ref{a2}). One indeed obtains excellent  agreement between the spectra
of the mesonic and baryonic states (See Fig. \ref{rho-delta}). The
values of $\sqrt{\la_M}$ and $\sqrt{\la_B}$ are nearly degenerate
as predicted by superconformal quantum mechanics.

There is, however, a problem with the $\rho/a_2$-$\Delta$
connection in that  half-integer twist is apparently required. For
the $\Delta $ trajectory the observed spectrum corresponds to
half-integer twist $2 +L_B+ \half$, which also implies
half-integer twist  for the mesons on the $\rho/a_2$ trajectory.
Although the spectra of this half-integer twist obtained with the
superconformal Hamiltonian operator \req{a1} correspond fully to
those obtained by LFHQCD (and experiment), the wave functions do
not;  they differ by a factor $x^{\half}$.   Related to this
problem is the fact that the supercharge $R_\la^\dagger$ does not
annihilate the $\rho$ wave function, but it formally leads to a
baryonic state with the same mass.  However, this state is
excluded as a physical state, since it would have the angular
momentum $L_B= -1$.

It should be noted that the semiclassical equations of light-front
holographic QCD and superconformal quantum mechanics are intended
to be  a zeroth order approximation to the complex problem of
bound states in QCD. We also emphasize that the quantum mechanical
supersymmetric relations derived here are not a consequence of a
supersymmetry of the underlying quark and gluon fields; they are
instead  a consequence of the superconformal-confining dynamics of
the semi-classical theory and the clustering inherent in
light-front  holographic QCD.

In this paper we have
concentrated on the consequences of superconformal algebra for 
the spectral properties of meson and baryons. Since the meson and baryon wave functions are also related,
there are  also interesting dynamical consequences;  {\it e.g.}, for  elastic and transition form factors.
The  $b_1$ wave function  is  predicted to be identical
to  the non-leading-twist wave function of the nucleon, which in
turn is related to the leading-twist wave function via
a parity transform -- see \cite{deTeramond:2014asa}; therefore, at low resolution
the form factors of the nucleon and the $b_1$ are related.   Another
dynamical consequence of the model is that for high resolution,  at large  momentum transfer when the baryon cluster is resolved into its individual constituents, the
twists of the superpartners are equal: the higher value of $L$ of
the meson,  $L_M = L_B + 1$, is compensated by the additional constituent in
the baryon. 

\vspace{30pt}

{\bf \Large Acknowledgements}

\vspace{20pt}

{ We thank Adi Armoni for helpful comments.} The work of SJB was
supported by the Department of Energy contract
DE--AC02--76SF00515.

\appendix

\section{Other Possible Evolution Operators \label{Op}}

Fubini and Rabinovici have discussed three  different ways of
constructing compact Hamiltonians from the superconformal algebra.
Some care should be taken, however, in transferring their
interpretation to our application. The emphasis
in~\cite{deAlfaro:1976je} and later
in~\cite{Akulov:1984uh,Fubini:1984hf} was on quantum mechanics as
a one dimensional field theory and the investigation of the vacuum
structure in this field theory. Therefore they only  the
case with a stable classical potential, implying $f > 0$,  was considered. In our
search for semiclassical bound-state equations, however, the
lowest state is a hadronic state. Furthermore, in the field
theoretical investigations of FR the dimensionless constant $f$ is
an arbitrary positive parameter, each value of $f$ representing a
different field theory with a different vacuum. In our
investigations, where the procedure of dAFF~\cite{deAlfaro:1976je}
and its extension by FR~~\cite{Fubini:1984hf} has been embedded in
LFHQCD  the dimensionless constant $f$ determines the angular
momentum and we are confined to the series of discrete values
representing the  orbital excitations.  Nevertheless, it is
informative to discuss the three different ways to construct the
compact Hamiltonian  representing hadronic bound states also from
our perspective.  For generality purposes we use in the appendices
for the dimensionful constants the symbols $w$ and $v$, which can
be positive or negative.  In our applications to meson and baryon
spectroscopy we are restricted $w  > 0$.

The simplest way to construct a Hamiltonian with discrete spectrum
in the frame of the superconformal  graded algebra is to apply
directly the method of dAFF~\cite{deAlfaro:1976je}. This yields
the Hamiltonian~\cite{Akulov:1984uh, Fubini:1984hf}, again in the
slightly generalized notation 
\beq G_0 = \{Q,Q^\dagger\} +w^2\, K.
\enq 
Both supersymmetry and dilatation symmetry are broken here.
The two components of the eigen-spinor of $G_0$ have different
spectra
 \beqa
  {(G_0)}_{11} \phi_1 &=& (4 n + 2 f +3)|w|\phi_1, \\
 {(G_0)}_{22} \phi_2 &=& (4 n + 2 f +1)|w|\phi_2,
\enqa and thus supersymmetry is broken from the onset for all
levels. This approach would yield a LF potential $U(\zeta) =w^2
\zeta^2$, without any additional constants which occur in LFHQCD
(See \ref{a5}, \ref{a3}, \ref{a4}), and which are
phenomenologically very important.

On the other hand, the {approach} where supersymmetry is conserved
by constructing a new Hamiltonian from the { spinor operator}
$R_w$, {a superposition of the supercharges $Q$ and $S$ within the
superalgebra~\cite{Fubini:1984hf}}, \beq  { R_w = Q + w S, } \enq
conserves supersymmetry for $f>\half$, since $R_w$ commutes with {
the evolution operator} \beq \label{G} G(w) = \{R_w,R_w^\dagger\}.
\enq Therefore $R_w|\phi_w\rangle$ is an eigenstate of $G_w$ {
with identical eigenvalue as the eigenstate $|\phi_w\rangle$}.

The spectra of $G(w)$ for  real {values} of $f$ and $w$ are: \beqa
\label{spec-gen}
E_1&=& (4n +2) |w| +2\, \left\vert f+\half \right\vert \,|w| + 2 (f-\half) \, w, \\
E_2&=& (4n +2) |w| +2\, \left\vert f-\half \right\vert \,|w| + 2
(f+\half) \, w , \enqa where $E_1$ { are eigenvalues of $G_{11}$
representing  mesonic states and $E_2$ the  eigenvalues of
$G_{22}$ for baryons}. For $w < 0$ { and $f > - \half$} the
spectra are independent of $f$: \beqa
E_1 &=&4(n+1)|w|,\\
E_2&=&4n |w| , \enqa and therefore cannot lead to angular
excitations of the corresponding LF Hamiltonians. For $f=-\half$
and $w > 0$, we have \beqa
E_1 &=& 4 n \,  w ,\\
E_2 &=& 4(n+1)\,   w . \enqa There exists no baryonic state \beq
|\phi\rangle = \left(\begin{array}{c} 0\\
\phi_B\end{array}\right), \enq with zero energy. The reason for
this seeming contradiction with the above mentioned commutation
relation, lies in the fact that the operator $R_w^\dagger$
annihilates the mesonic state (See \req{pi-shift}).

\section{Transformation Operators and Quantum Mechanical Evolution \label{Oq}}

The generalized hypercharge  $R$ has the commutation relations
\beqa \label{com1}
[G(w), R_v]   &=&  -2(w-v)\,  R_{-w} , \\
\label{com2}
{[G(w) ,R^\dagger_v ]}    &=&  2(w-v)\,  R^\dagger_{-w}, \enqa
with the new Hamiltonian $G(w) =\{R_w, R_w^\dagger\}$.

For $v=w$ the commutator vanishes, therefore { if $\vert \phi
\rangle$ is an eigenstate of $G$,  also $R_w\, \vert \phi \rangle$
is an eigenstate}  with the same eigenvalue. Therefore { the
spinor supercharge $R$} transforms the baryonic superpartner with
angular momentum $L_B$, into the mesonic one with angular momentum
$L_M=L_B+1$. The operator $R_w^\dagger$ { acts}  in the opposite
direction.

For $v=-w$, however, we have the typical commutation behavior of a
raising and lowering operator, respectively: \beqa
 [G(w),R_{-w}]   &=& - 4 w \, R_{-w}, \\
{ [G(w),R_{-w}^\dagger]} &=& 4 w \, R_{-w}^\dagger . \enqa That
is, if  $|\phi \rangle$ is eigenfunction of $G$ with eigenvalue
$E$, then $R_{-w} \, |\phi \rangle$ is eigenstate with the energy
$E+4 w$. { This means that} a baryonic state with angular momentum
$L_B$ and radial excitation $n$ is transformed into a mesonic
state with angular momentum $L_M=L_B+1$ and radial excitation
$n+1$, which has the same energy as the baryonic state with
angular momentum $L_B$ and radial excitation $n+1$.

There is also a bosonic raising operator, that is,  a raising
operator which does not change fermion number. It is composed of
the bosonic operators of the graded algebra. Generalizing again
slightly the operators introduced by  FR in
Ref.~\cite{Fubini:1984hf} \beq  \label{Lv} L_v = H + v^2 \,K + 2 i
\,v\,D,
 \enq
 one obtains from the algebra \req{susy-extended}
the commutation relations \beqa
[G(w),L_w] &=&  4 w \,L_w , \\
{[G(w),L_{-w}]} &=& -4 w \,L_w . \enqa {These relations imply
that} also $L_w$ is a raising operator, which transforms a baryon
with $L_B,\, n$ into a baryon with $L_B,\, n+1$ and the same with
the mesons. Since it is composed of operators of the conformal
group, it  can also be applied to the lowest mesonic state,
although there is no supersymmetric partner.

\begin{figure}

\begin{center}
\setlength{\unitlength}{1.2mm}

\begin{picture}(90,90)(-50,0)
{\large \put(-40,0){Baryons, $ L_B$} \put(15,0){Mesons,
$L_M=L_B+1$ } \put(-40,10){\vector(0,1){68}} \put(-40,79){$n$}
\put(-40,10){\line(1,0){2}} \put(-43,10){0}
\put(-40,30){\line(1,0){2}} \put(-43,30){1}
\put(-40,50){\line(1,0){2}} \put(-43,50){2}
\put(-40,70){\line(1,0){2}} \put(-43,70){3}

\put(-30,10){\circle*{3}} \put(30,10){\circle{3}}
\put(-30,32){\circle*{3}} \put(30,32){\circle{3}}

\put(0,12){$R_w$} \put(18,23){$R_{-w}^\dagger$}
\put(-28,11){\vector(1,0){56}} \put(-28,12){\vector(3,1){56}}
\put(0,4){$R_w^\dagger$} \put(28,9){\vector(-1,0){56}}
\put(-28,30){\vector(3,-1){56}} \put(-23,23){$R_{-w}$}
\put(-31,12){\vector(0,1){17}} \put(-29,29){\vector(0,-1){17}}
\put(-36,18){$L_w$} \put(-28,18){$L_{-w}$}
\put(31,12){\vector(0,1){17}} \put(29,29){\vector(0,-1){17}}
\put(24,18){$L_w$} \put(32,18){$L_{-w}$}

\put(-30,54){\circle*{3}} \put(30,54){\circle{3}}

\put(0,34){$R_w$} \put(18,45){$R_{-w}^\dagger$}
\put(-28,33){\vector(1,0){56}} \put(-28,34){\vector(3,1){56}}
\put(0,27){$R_w^\dagger$} \put(28,31){\vector(-1,0){56}}
\put(-28,52){\vector(3,-1){56}} \put(-23,45){$R_{-w}$}
\put(-31,34){\vector(0,1){17}} \put(-29,51){\vector(0,-1){17}}
\put(-36,40){$L_w$} \put(-28,40){$L_{-w}$}
\put(31,34){\vector(0,1){17}} \put(29,51){\vector(0,-1){17}}
\put(24,40){$L_w$} \put(32,40){$L_{-w}$}

\put(-28,55){\vector(1,0){56}} \put(-28,56){\vector(3,1){56}}
\put(28,53){\vector(-1,0){56}} \put(-28,74){\vector(3,-1){56}}
\put(-31,56){\vector(0,1){17}} \put(-29,73){\vector(0,-1){17}}
\put(31,56){\vector(0,1){17}} \put(29,73){\vector(0,-1){17}}

}
\end{picture}
\end{center}
\caption{\label{SCO} \small Radial excitations and transformations
by elements of the superconformal algebra for a baryon-meson
system with a given $f-\half =L_B \geq 0$.}
\end{figure}
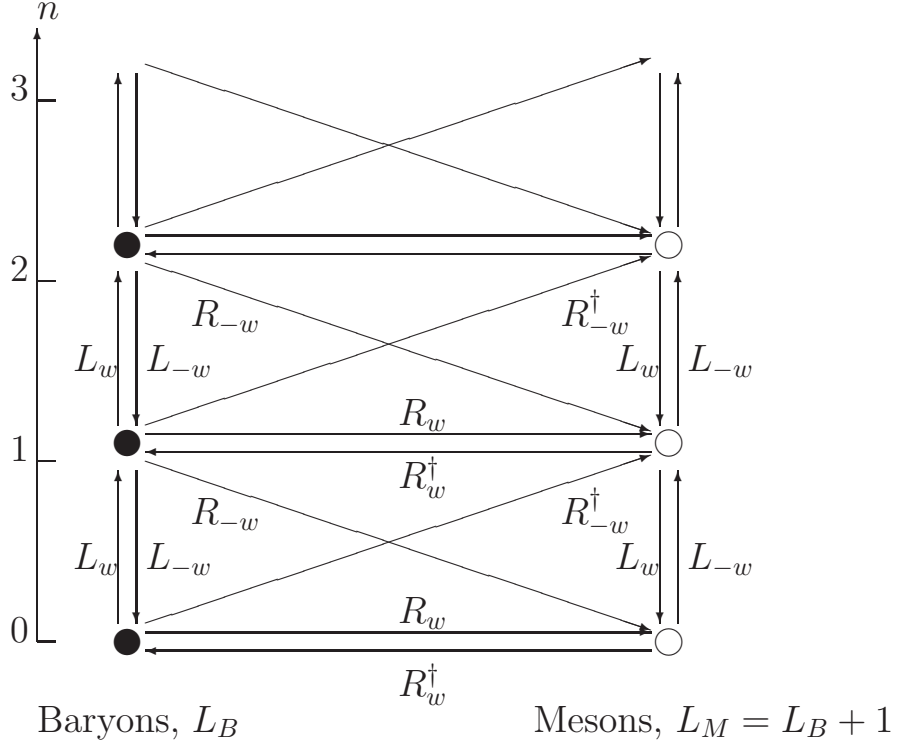

Since the  hypercharges $R_w$ change the angular momentum by one
unit, it is tempting to look for an operator which also leads to
angular excitations. Such an operator which increases the angular
momentum by one unit  is easily constructed and has the form:
  \beq \label{La}
\La_w =\{Q,\psi\} +w\{S^\dagger, S\}+ \frac{1}{\zeta} \psi^\dagger
\psi. \enq If $|\phi \rangle_L$ is an eigenstate to the Hamiltonian
operator $G_L$ constructed with $f=L+\half$, then $|\phi
\rangle_{L+1} = \La_w\, |\phi \rangle_L$ is eigenstate to
$G_{L+1}$, constructed with $f=L+1+ \half$. This operator $\La$
is, however, not an element of the superconformal algebra. The
action of the different operators in the baryon-meson system is
illustrated in Fig. \ref{SCO}.

\subsection{Quantum-Mechanical Evolution}

In this paper, as in \cite{deTeramond:2014asa}, we have
concentrated on algebraic aspects and its consequences for the
spectra. We now briefly discuss the  quantum-mechanical  time
evolution. The Hamiltonian of unbroken superconformal quantum
mechanics, $H$ Eq. \req{Ham},  is the translation operator for the
time variable $t$ \beq \label{SE} i\,\frac{d}{dt} \vert
\phi\rangle = H\,   \vert \phi\rangle. \enq The
quantum-mechanical evolution of the operator \req{Gwu} \beq G=2
H+2  w^2\,K +2 \la  \,( f \mbox{\bf I} - B), \enq follows from the
action of the generators $H$ and $K$ on  the state $\vert \phi
\rangle$. We have (See Appendix C in Ref.~\cite{Brodsky:2014yha}),
\beqa
e^{-i \,H\,\ep}\vert \phi(t) \rangle&=&\vert \phi(t) \rangle + \frac{d}{dt}\vert \phi(t) \rangle \ep + O(\ep^2),\\
e^{-i \,K\,\ep}\vert \phi(t) \rangle&=&\vert \phi(t) \rangle +
\frac{d}{dt}\vert \phi(t) \rangle \ep \,t^2 + O(\ep^2). \enqa
There follows \beq
 G \, \vert \phi(\tau) \rangle =\left(i \frac{d}{d\tau}
+2 \la  \,( f \mbox{\bf I} - B)\right) \vert \phi(\ta) \rangle,
\enq where the new evolution parameter $\tau$ is related to $t$ in
\req{SE} by \beq d \tau = \frac{dt}{2(1 + \la t^2)}, \enq as in
dAFF~\cite{deAlfaro:1976je}. From the eigenvalue equation $G\,
\vert \phi_E \rangle = E\, \vert \phi_E \rangle$ follows the
stationary state solution \beq
 \vert \phi_E(\tau) \rangle =  \vert \phi_E(0) \rangle  e^{-i\left((E \mbox{\bf \footnotesize I} - 2 \la \,( f \mbox{\bf \footnotesize I} - B) \right)\tau}.
 \enq

\subsection{Operators in Matrix Form}

It is sometimes convenient to work with a special matrix
representation of the superconformal algebra. For convenience we
give here an explicit realization in the Schr\"odinger picture.
Define \beq q =-\frac{d}{dx} + \frac{f}{x}, \quad \quad q^\dagger
= \frac{d}{dx}  + \frac{f}{x}. \enq Then we can write the spinor
operators $Q$ and $S$ as \beq Q = \left(\begin{array}{cc}
0&q\\
0&0\\
\end{array}
\right) ,\quad \quad Q^\dagger=\left(\begin{array}{cc}
0&0\\
q^\dagger&0\\
\end{array}
\right), \enq and \beq S = \left(\begin{array}{cc}
0& x\\
0&0
\end{array}
\right), \quad \quad S^\dagger=\left(\begin{array}{cc}
0&0\\
x&0\\
\end{array}
\right). \enq

The Hamiltonian  $H = \half \{Q,Q^\dagger\}$ in matrix form is
\beq 2H  =\left(\begin{array}{cc}
q\,q^\dagger&0\\
0&q^\dagger\,q
\end{array}
\right)= \left(\begin{array}{cc}
- \frac{d^2}{d x^2}+\frac{f(f+1)}{x^2}&0\\
0& - \frac{d^2}{d x^2} + \frac{f(f-1)}{x^2}
\end{array}
\right) . \enq

The Hamiltonian $G =\{R_w, R_w^\dagger\}$ is \beq G
=\left(\begin{array}{cc} - \frac{d^2}{d x^2} + w^2 \,x^2+ 2 w\, f
- w+ \frac{4 (f+\half)^2 - 1}{4 x^2}&\hspace{-1cm} 0
\\
0&\hspace{-1cm} - \frac{d^2}{d x^2} + w^2 \, x^2 + 2 w\, f  + w +
\frac{4 (f-\half)^2 -1}{4 x^2}
\end{array}\right),
\enq where \beq ~~~~~ R_w   =\left(\begin{array}{cc}
0&-\frac{d}{dx}+ \frac{f}{x} + w\,x\\
0&0
\end{array}\right),
\qquad \enq and \beq
 R_w^\dagger=\left(\begin{array}{cc}
0&0
\\
\frac{d}{dx}+ \frac{f}{x} + w\,x&0
\end{array}\right).
\enq

The operator $L_w$ \req{Lv} is \beq L_w= H +\half\left(w^2\,x^2 -
\frac{d}{dx}\,x -x\,\frac{d}{dx}\right)  {\bf I}, \enq and its
adjoint \beq
 L_w^\dagger= H +\half\left(w^2\,x^2 + \frac{d}{dx}\,x +x\,\frac{d}{dx}\right) {\bf I}.
\enq

Finally, the orbital raising operator \req{La} is \beq \La_w
=\left(\begin{array}{cc}
-\frac{d}{dx} +\frac{f+1}{x} +w \,x^2&0\\
0&-\frac{d}{dx} +\frac{f}{x} +w \,x^2
\end{array}\right).
\enq

In this matrix form the upper component of the state
$|\phi\rangle$ is the meson, the lower one the baryon \beq \vert
\phi\rangle = \left(\begin{array}{c} \phi_M \\  \phi_B
\end{array} \right).
\enq 
Thus the effective baryon number} operator $N_B =
\half (1 - [\psi^\dagger, \psi])$ is in matrix form: \beq N_B =
\left(\begin{array}{cc}
0&0\\
0&1\\
\end{array}
\right). \enq 

It is easy to check that the state containing the pion, that is
the eigenstate of \req{a1} with $f=-\half$, namely \beq \phi_\pi =
\frac{1}{N} \sqrt{x\,} e^{- w\,x ^2/2}, \enq has no supersymmetric
partner, since \beq \label{pi-shift} R_w ^\dagger \,| \phi\rangle
= \left(\begin{array}{c} 0 \\ (q^\dagger +w\,x)  \phi_\pi
\end{array} \right) =  \left(\begin{array}{c} 0 \\ 0  \end{array}
\right) . \enq Likewise, one checks that the state containing the
$\rho$-meson, where $f=0$, with the wave function \beq \phi_\rho=
\frac{1}{N} x \, e^{- w\,x ^2/2}, \enq has formally a
superpartner, but with negative angular momentum $L_B=-1$. Indeed
\beq \label{rho-shift} R_w ^\dagger \,| \phi\rangle =
\left(\begin{array}{c} 0 \\ (q^\dagger +w\,x)  \phi_\rho
\end{array} \right) =  \left(\begin{array}{c} 0\\ \phi_\rho
\end{array} \right).
\enq



\end{document}

%% file: definitions.tex
\def\Dslash{\raise.15ex\hbox{/}\kern-.7em D}
\def\Pslash{\raise.15ex\hbox{/}\kern-.7em P}

\newcommand{\beq}{\begin{equation}}
\newcommand{\enq}{\end{equation}}
\newcommand{\beqa}{\begin{eqnarray}}
\newcommand{\beqast}{\begin{eqnarray*}}
\newcommand{\enqa}{\end{eqnarray}}
\newcommand{\enqast}{\end{eqnarray*}}

\newcommand{\req}[1]{(\ref{#1})}

\newcommand{\half}{{\frac{1}{2}}}
 \newcommand{\threehalf}{{\frac{3}{2}}}

\newcommand{\bec}{\begin{center}}
\newcommand{\enc}{\end{center}}
\newcommand{\beqo}{\begin{quote}}
\newcommand{\enqo}{\end{quote}}

\newcommand{\cL}{{\cal L}}

\newcommand{\cN}{{\cal N}}

\newcommand{\al}{\alpha}

\newcommand{\ga}{\gamma}

\newcommand{\ep}{\epsilon}

\newcommand{\la}{\lambda}

\newcommand{\si}{\sigma}

\newcommand{\ta}{\tau}

\newcommand{\vp}{\varphi}

\newcommand{\La}{\Lambda}